\documentstyle[12pt]{article}

\begin{document}
\date{}
\large

\title{Time dependent neutrino billiards}

\author{D.U.Matrasulov, Z.A.Sobirov, Sh.Ataev, H.Yusupov\\
Heat Physics Department of the Uzbek Academy of Sciences,\\
28 Katartal St.,700135 Tashkent, Uzbekistan}
\maketitle

    \begin{abstract}
    Quantum dynamica of a massless Dirac particle in time-dependent 1D box and circular billiard with time-dependent radius is studied.
    An exact analytical wave functions and eigenvalues are obtained for the case of linear time-dependence of the boundary position.
    \end{abstract}

Neutrino billiard was discussed first by Berry and Mondragon in the context of time-reversal symmetry breaking and quantum chaos in relativistic systems\cite{berry}.
Recently 2D, massless Dirac equation has attracted much attention in the context of graphene physics.
Graphene is a planar monolayer of carbon atoms tightly packed into a two-dimensional honeycomb lattice.
The quasiparticle motion in graphene in the low-energy regime is described by zero-mass Dirac equation with different(bulk and nanoribbon graphenes) boundary conditions \cite{been,katsn,brey}.

Also, so-called graphene billiard has become attractive topic \cite{burg} in the context of graphene physics.

 In this work we treat the problem of neutrino billiard with moving boundaries.
Besides its relevance to graphene billiard such system is of importace in the context of relativistic quantum Fermi acceleration, too.
Studying of time-dependent quantum requires solving of the two-dimensional Schr\"odinger or Dirac equations with time-dependent boundary conditions. We note that quantum mechanical wave equations with time-dependent boundary conditions
 have been subject of  extensive study during past three decades \cite{doescher}-\cite{sch91}.
   The early treatment of the one-dimensional Schr\"odinger   equation with non-stationary
    boundary conditions dates back to Doescher \cite{doescher} who explored  quantum
    dynamics of a particle in square well with moving walls. Later  Munier\cite{mun81} and Pinder
    \cite{pinder}  studied this problem for some special cases.
    More  comprehensive and mathematically strict treatment of the Schr\"odinger equation with time-dependent   boundary conditions
    has been done in a series of papers by Makowski et al \cite{mak91}-\cite{mak923}. Seba has studied this problem in the context of quantum Fermi acceleration \cite{seb90}. The case of time-periodic boundary condition is treated by Scheiniger and Kleber \cite{sch91}. Three-dimensional case was treated by Yuce for the case of spherical boundaries \cite{yuce}
    Despite the progress made in the study of Schr\"odinger equation with time-dependent
    boundary conditions,
    conditions, relativistic extension of this problem for the case of Dirac equation with moving boundary conditions has not yet been
    considered.
    In this work  we address the problem of  massles Dirac particle in time-dependent confined geometries considering most simplest cases, time-dependent 1D box and time-dependent circular billiard.
   For the case of special time-dependence of the boudnary conditions we obtain analytically eigenfucntions and eigenvalues
   of the Dirac equation for massless particle.
\section{Time-dependent one dimensional box}
Consider the following time-dependent Dirac system ($\hbar = c=1$)
\begin{equation}\label{dir1}
i\frac{\partial \Psi}{\partial t}= (\alpha p+m\beta)\Psi,
\end{equation}
with $\Psi(t,x)=(\Psi_1(t,x),\Psi_2(t,x))^T$ being two-component spinors,
in the domain
$$D=\{(t,x), \;0<t<T,\; 0<x<L(t)\},
$$
where the right boundary $L(t)$, is time-dependent and
$$p=-i\frac{\partial}{\partial x}$$ and
$$\alpha =
\left(\begin{array}{cc} 0 & -i\\i & 0   \end{array}, \right), \
\beta = \left(\begin{array}{cc} 1 & 0\\0 & -1   \end{array},
\right)$$
are the Dirac matrices.
for which the boundary conditions are given as
\begin{equation}\label{gr1} \Psi_1(t,0)=0,\ \ \Psi_1(t, L(t))=0, \ \ 0\leq
t\leq T \label{bound}
\end{equation}

The Dirac equation for (fixed) boundary conditions, given in the
one-dimensional box, ( $L(t)=const$)  was treated in refs.
\cite{Alonso,Alonso1} where  the solution of the Dirac equation
$$(\alpha p+ mc^2\beta)\Psi=E\Psi$$
is obtained as
$$\Psi_n(x) = A_n \left( \begin{array}{c} \sin(k_nx)\\-\frac{ck_n}{E_n+mc^2}\cos(k_nx)
\end{array}\right), $$
with $A_n$ being the normalization constant and  $k_n=n\pi /L, n=1,2, ...$,
$E_n=[k_n^2+(mc^2)^2]^{1/2}$.

To solve the eq. \ref{dir1} we will restrict ourselves by
considering the massless case, i.e., we assume that $m=0.$ The
eq.\ref{dir1}  with the boundary conditions \ref{bound} cannot be
solved until time-dependent boundary conditions are not replaced
with fixed ones. To do this we use in the eq.\ref{dir1} the
following substitution: $y=\frac{x}{L(t)}$ that reduced the domain
$D$ to $D_1=\{ (t,x), 0<t<T,0<y<1\}$.

In this case the eq. (\ref{dir1}) can be rewritten as

\begin{equation}\label{dir2}
\left\{\begin{array}{l} i\frac{\partial \Psi_1 }{\partial t}=
L^{-1}\frac{\partial \Psi_2 }{\partial
 y}+i\dot{L}L^{-1}y\frac{\partial \Psi_1 }{\partial y},\\
 i\frac{\partial \Psi_2 }{\partial t}= -L^{-1}\frac{\partial \Psi_1 }{\partial
 y}+i\dot{L}L^{-1}y\frac{\partial \Psi_2 }{\partial
 y}\end{array}\right. ,
\label{dir4}
\end{equation}
where $\dot{L}=\frac{dL}{dt}$, and the conudary conditions are given as
\begin{equation} \label{gr2} \Psi_1(t,0)=0,\ \ \Psi_1(t, 1))=0, \end{equation}

Time and coordinate variables in the eq.\ref{dir3} cannot be
separated for arbitrary time-dependence of $L(t)$. The only case
for which variables can be separated is $L(t)=at+b$.

Introducing new time variable
 $$\tau =\int\limits_0^t\frac{ds}{L(s)}=\frac{1}{a}\ln\left(\frac{at+b}{b}\right),$$
and using the substitutions $\Psi_1(\tau ,y) =
e^{-i\lambda\tau}f(y)$, $\Psi_1(\tau ,y) = e^{-i\lambda\tau}g(y)$
we get from the eq. (\ref{dir2})
\begin{equation}\label{dir3}
\left\{\begin{array}{l}\frac{d g}{d y}+i a y \frac{d
f}{d y}=\lambda f,\\
- \frac{d f}{d y}+i a y \frac{d g}{d y}=\lambda g.
\end{array}\right.
\end{equation}
This system can be reduced to the second-order equation as
\begin{equation}
\left\{(1-a^2y^2)\frac{d^2}{dy^2}-2ay(i\lambda +a)\frac{d}{dy}-i
a\lambda \right\}f=-\lambda^2f. \label{hg1}
\end{equation}

Using the substitution  $$z= \frac{1-ay}{2},$$ the eq. (\ref{hg1})
for $|a|<1$ can be reduced to the following equation
$$z(1-z)f^{\prime\prime}+(c-(a+b+1)z)f^{\prime}-abf=0.$$
whose solutions are hypergeometric functions.

In our case the hypergeomtric series can be written as
$$F_1(z)=F(\alpha_1, \beta_1, \gamma_1, z)=\sum_{k=0}^{+\infty}\frac{(\beta_1)_k}{k! }z^k = (1-z)^{\beta_1}.$$
Second  solution can be found as
$$F_2(z)=z^{\beta_1}.$$

Then the general solution of the eq. (\ref{hg1}) can be written
as
\begin{equation}\label{solhg1} f(y)=A\left(\frac{1-ay}{2}\right)^{-i\lambda/a}+
B\left(\frac{1+ay}{2}\right)^{-i\lambda/a},   \end{equation} where
$A^i=\exp(i \ln(a))$

Then from the boundary conditions given by the eq. (\ref{gr2}) we
get
\begin{equation}\label{L1} A=-B, \;\ \ \left(\frac{1-a}{2}\right)^{-i\lambda/a}
-\left(\frac{1+a}{2}\right)^{-i\lambda/a}=0.  \end{equation}
Solving this equation we obtain the eigenvalues as
\begin{equation}\lambda_n={2a\pi
n}\left|\ln|1-a|-\ln|1+a|\right|^{-1}.\label{eigenvalues1}\end{equation}

Corresponding eigenfunctions are

\begin{equation}\label{sol3}\begin{array}{l}
f_n(y)=M\left(\left({1-ay}\right)^{-i\lambda_n/a}-
\left({1+ay}\right)^{-i\lambda_n/a}\right),\\
g_n(y)=\frac{M}{i\lambda_n -
a}\left(\left({1-ay}\right)^{-i\lambda_n/a+1}+
\left({1+ay}\right)^{-i\lambda_n/a+1}\right)-\\
\ \ \ \ \ \ \ \ \ \ \
iayM\left(\left({1-ay}\right)^{-i\lambda_n/a}-
\left({1+ay}\right)^{-i\lambda_n/a}\right)
\end{array}
\end{equation}

We note that
for $a\to 0$ the eigenvalues coincide with
$\lambda_n\to\pi n$,and eigenfucntions become
$$\begin{array}{c}
f_n(y)\to A sin(\pi n y),\\
g_n(y)\to -\frac{A}{\pi n}cos(\pi n y).\end{array}
$$

Then the complete set of solutions of the eq.
 (\ref{dir1}) for linearly moving boundaries can be written as

\begin{equation}\label{sol1}\begin{array}{l}
\Psi_{1n}(t,x)=M\exp\left[-\frac{i\lambda_n}{a}\ln\left(\frac{at+b}{b}\right)
\right]\left(\left({1-\frac{ax}{at+b}}\right)^{-i\lambda_n/a}-
\left({1+\frac{ax}{at+b}}\right)^{-i\lambda_n/a}\right),\\
\Psi_{2n}(t,x)=M\exp\left[-\frac{i\lambda_n}{a}\ln\left(\frac{at+b}{b}\right)
\right]\cdot\\
\ \ \ \ \ \ \ \ \ \ \ \ \ \ \cdot\left\{\frac{1}{i\lambda_n -
a}\left(\left({1-\frac{ax}{at+b}}\right)^{-i\lambda_n/a+1}+
\left({1+\frac{ax}{at+b}}\right)^{-i\lambda_n/a+1}\right)-\right.\\
\ \ \ \ \ \ \ \ \ \ \ \ \ \ \left.
-iay\left(\left({1-\frac{ax}{at+b}}\right)^{-i\lambda_n/a}-
\left({1+\frac{ax}{at+b}}\right)^{-i\lambda_n/a}\right)\right\}.
\end{array}
\end{equation}

    Thus we have treated the Dirac equation for massless particle in one-dimensional
    infinite square well with a time-dependent wall.

\section{Time-dependent circular billiard.}

Consider the Dirac equation with the boundary conditins given at the
circle with time-dependent radius $x^2+y^2<r_0^2(t)$.
Radial equation can be written as
\begin{equation}\label{dirR}
\begin{array}{c}
i\frac{\partial P(r,t)}{\partial t}=\frac{\partial
Q(r,t)}{\partial r} -\frac{k}{r}Q(r,t),\\

i\frac{\partial Q(r,t)}{\partial t}=-\frac{\partial
P(r,t)}{\partial r} -\frac{k}{r}P(r,t),
\end{array}
\end{equation}
whereü $k$ is an integer number.
Boundary condition is given by
\begin{equation}
P(r_0(t),t)=0.\label{bcR}
\end{equation}

Using substitution
$$
y=\frac{r}{r_0(t)}, \ \ \tau=\int\limits_{0}^{t}\frac{ds}{r_0(s)},
$$
the eqs.(\ref{dirR}) and (\ref{bcR})can be rewritten as
\begin{equation}\label{dirR2}
\begin{array}{c}
 i\frac{\partial P(y,\tau)}{\partial \tau}=i\dot r_0 y
\frac{\partial P(y,\tau)}{\partial y} +\frac{\partial
Q(y,\tau)}{\partial y}-\frac{k}{y} Q(y,\tau),\\

i\frac{\partial Q(y,\tau)}{\partial \tau}=i\dot r_0 y
\frac{\partial Q(y,\tau)}{\partial y} -\frac{\partial
P(y,\tau)}{\partial y}-\frac{k}{y} P(y,\tau),
\end{array}
\end{equation}
where the boundary condition is time-independent now and given as
\begin{equation}
P(1,t)=0.\label{bcR2}
\end{equation}

Separating time and coordinate variables in the eq.(\ref{dirR2})   for $\dot r_0(t)=0$ and $r_0(t)=at+b$ and reducing the obtaibned first-order system into second order equation
we have

\begin{equation}\label{eq1}
y^2(a^{2}y^2-1)\frac{d^2 f}{d y^2} +2a(a+i\lambda)y^3\frac{d f}{d
y} +[(ia\lambda-\lambda^2)y^2+k(k+1)]f=0
\end{equation}

Solution of this equation can be written in terms of (regular at $y=0$)hypergeometric function as
$$
f_{\lambda}(y)=y^{\gamma} F\left(\alpha, \alpha+\frac{1}{2},
\gamma, a^2y^2\right),
$$
where
$$\alpha=\frac{1}{2}\left|k+\frac{1}{2}\right|+\frac{i\lambda}{2m}+\frac{1}{4}\ \ \gamma=\left|k+\frac{1}{2}\right|+1.$$

For secnd component we have
$$
g_{\lambda}(y)=(\lambda-i(1-k)a)y^k\int\limits_0^y\frac{f_{\lambda}(s)ds}{s^k}-iayf_{\lambda}(y).
$$

The eigenvalues $\lambda_n$ can be found from the  boundary condition:
 \begin{equation}\label{eigen}
 F\left(\frac{1}{2}\left|k+\frac{1}{2}\right|+\frac{i\lambda}{2m}+\frac{1}{4}, \frac{1}{2}\left|k+\frac{1}{2}\right|+\frac{i\lambda}{2m}+\frac{3}{4}, \gamma,
 a^2\right)=0.
\label{eqq}
 \end{equation}

For  $k=0$ the solution of the eq.\ref{eqq} can be found analytically:
$$
\lambda_n=2\pi n\ a / \ln \left(\frac{1+a}{1-a}\right)
$$

The solutions of the eqs.(\ref{dirR}) and (\ref{bcR}) for  $k=0$  can be written as
$$
P_n(r,t)=N\exp\left(-\frac{i\lambda_n}{2a}\ln\left(\frac{at+b}{b}\right)\right)
f_{\lambda_n}\left(\frac{r}{r_0(t)}\right),
$$

$$
Q_n(r,t)=N\exp\left(-\frac{i\lambda_n}{2a}\ln\left(\frac{at+b}{b}\right)\right)
g_{\lambda_n}\left(\frac{r}{r_0(t)}\right),
$$
where $N$ is the normalization constant. Thus we have obtained analytically the solution of the massless
 Dirac equation for time-dependent 1D-box and circular billiard.
The circle is considered as monotonically expanding(contracting) with constant velocity. For general case (e.g., for breathing circle etc.) the solution can be obtained numerically.
The above system is of importance because of its relevance to quantum Fermi acceleration in relativistic systems and particle transport graphene as in real situation the boundaries on such system is not strictly fixed.
Another important aspect of the above treated problem is its relevance to Dynamical (fermionic) Casimir effect.
In this context the extension of the above problem to the case of non-integrable billiard geometries and time-periodic  boundaries could be important. Currently such studies are in progress.

\end{document}